\documentclass[twocolumn,showpacs,showkeys,preprintnumbers,amsmath,amssymb]{revtex4-1}

\usepackage{graphicx}
\usepackage{dcolumn}
\usepackage{bm}

\begin{document}

\preprint{Submitted to Ann.\ Geophys.}  

\title[Auroral vortex street]{Auroral vortex street formed by \\ the magnetosphere-ionosphere coupling instability}

\author{Yasutaka Hiraki}
 \email{hiraki.yasutaka@nipr.ac.jp}
\affiliation{
National Institute of Polar Research, Midori-cho 10-3, Tachikawa, Tokyo, 190-8518, Japan.\\
}


\date{\today}

\begin{abstract}
By performing three-dimensional nonlinear MHD simulations including Alfv$\acute{\rm e}$n eigenmode perturbations most unstable to the ionospheric feedback effects, we reproduced the auroral vortex street that often appears just before substorm onset. We found that an initially placed arc splits, intensifies, and rapidly deforms into a vortex street. We also found that there is a critical convection electric field for growth of the Alfv$\acute{\rm e}$n eigenmodes. The vortex street is shown to be a consequence of coupling between the magnetospheric Alfv$\acute{\rm e}$n waves carrying field-aligned currents and the ionospheric density waves driven by Pedersen/Hall currents. 
\end{abstract}

\maketitle

\section{Introduction}
The problem of substorm onset has occupied the literature on solar-terrestrial physics for the past fifty years since Akasofu [1964], and the current understanding, as established by high-resolution ground and satellite optical observations [Donovan et al., 2006; Sakaguchi et al., 2009; Henderson, 2009], is that the auroral arc initially deforms into a vortex street on the scale of 30--70 km. It is clearly observed that the vortex street originates from a preexisting or newly produced arc that intensifies in $\approx1$ min and expands poleward over the course of 2 to 3 min [Lyons et al., 2002; Mende et al., 2009]. 

The vortex street has been interpreted in terms of instabilities in the plasma sheet, e.g., shear flow and ballooning instabilities [Voronkov et al., 1999]. However, this interpretation is only supported by the faint expectation that factors affecting strong magnetic or pressure fluctuations come from the external domain. On the other hand, multiple satellite observations [Ohtani et al., 2002] have suggested that such a situation is not necessary, and there is an alternative means of arc intensification. A simple scenario seems to be that an arc lying on a local field line becomes destabilized through changes in the global conditions, leading to connection to magnetotail plasma instabilities [c.f., Haerendel, 2010; Henderson, 2009]. 

If the scenario is limited to one explaining only deformation of the arc and not its poleward expansion, it can be viewed as a nonlinear evolution of shear Alfv$\acute{\rm e}$n waves in the magnetosphere-ionosphere (MI) coupling with non-uniform active field lines. A three-dimensional simulation of feedback instability in the system indicated that the waves induce strong magnetic and flow shears to produce vortex structures around the magnetic equator [Watanabe, 2010]. Various linear eigenmodes from low-frequency field line resonances to high-frequency ionospheric Alfv$\acute{\rm e}$n resonances have been shown to become destabilized in a dipole magnetic field [Hiraki and Watanabe, 2011; hereafter, HW]. These predictions are partially supported by evidence of the enhancement of the convection flow before substorm onset [Bristow and Jensen, 2007]. 

In this study, we performed three-dimensional simulations of shear Alfv$\acute{\rm e}$n waves in a full field line system with MI coupling, including an east-west aligned arc. We report new results: i) the initial arc splits and quickly deforms into a vortex street, ii) there is a critical convection electric field for its growth, and iii) the relationship is derived between the vortex size and the extent of intensification. Unlike the previous studies on arc evolution starting from arbitrary setups [Lysak and Song, 2008; Streltsov et al., 2012], we solve equations describing the nonlinear evolution of the most unstable Alfv$\acute{\rm e}$n eigenmode perturbations intrinsic in the full field line and show that arc deformation is a consequence of their growth. Starting from the simplest analysis, we dropped processes related to sharp $v_{\rm A}$ cavities, two fluid effects, and field-aligned electric fields. Note that processes in the high-$\beta$ plasma sheet are beyond the range of an approach based on magnetic perturbations.

\section{Model Description}
In order to elucidate the physics involved in auroral structures, nonlinear evolution of shear Alfv$\acute{\rm e}$n waves propagating along the dipole magnetic field $\mbox{\boldmath $B$}_0$ can be modeled by using two-field reduced MHD equations [e.g., Chmyrev et al., 1988; Lysak and Song, 2008]. The waves slightly slip ($\Omega/k_\perp \ll v_{\rm A}$) through the feedback coupling to density waves at the ionosphere. The system of interest is a field line of $L \approx 8.5$ with a length of $l\approx 7\times 10^4$ km and at a latitude of 70$^\circ$, where auroral arcs develop; note that it corresponds to the lower latitude in the tail magnetic field geometry. The field line position $s$ is defined as $s=0$ at the ionosphere and $s=l$ at the magnetic equator. We set a local flux tube: a square of ($l_\perp \times l_\perp$) with $l_\perp = 10^{-3} l \approx 70$ km at $s=0$, a rectangle of ($h_\nu l_\perp \times h_\varphi l_\perp$) at $s$, and ($\approx 3300$ km $\times$ $\approx 1700$ km) at $s=l$ using dipole metrics $h_\nu(s)$ and $h_\varphi(s)$ with $B_0(s) = 1/h_\nu h_\varphi$ [HW, 2011]. 

The electric field $\mbox{\boldmath $E$}$ is partitioned into a background convective part $\mbox{\boldmath $E$}_0$ ($\perp \mbox{\boldmath $B$}_0$) and the Alfv$\acute{\rm e}$nic perturbation $\mbox{\boldmath $E$}_1 = B_0 \mbox{\boldmath $\nabla $}_\perp \phi$. The magnetic perturbation is expressed as $\mbox{\boldmath $B$}_1 = \mbox{\boldmath $\nabla $}_\perp \psi \times \mbox{\boldmath $B$}_0$. The equations at $0<s \le l$ are written as 
\begin{eqnarray}
 \displaystyle && \partial_t \omega + \mbox{\boldmath $v$}_\perp \cdot \mbox{\boldmath $\nabla$}_\perp \omega = v_{\rm A}^2 \nabla_\parallel j_\parallel  \\
 && \partial_t \psi + \mbox{\boldmath $v$}_0 \cdot \mbox{\boldmath $\nabla$}_\perp \psi + \frac{1}{B_0} \nabla_\parallel B_0 \phi = -\eta j_\parallel. 
\end{eqnarray}
The convective drift velocity $\mbox{\boldmath $v$}_0=\mbox{\boldmath $E$}_0 \times \mbox{\boldmath $B$}_0/B_0^2$ is set so that $E_0$ satisfies the equi-potential condition, while $\mbox{\boldmath $v$}_\perp = \mbox{\boldmath $v$}_0 + \mbox{\boldmath $v$}_1(\mbox{\boldmath $E$}_1)$, vorticity $\omega = \nabla_\perp^2 \phi$, field-aligned current $j_\parallel = - \nabla_\perp^2 \psi$, and $\nabla_\parallel = \partial_s + \mbox{\boldmath $b$}_0 \cdot \mbox{\boldmath $\nabla $}_\perp \times \mbox{\boldmath $\nabla $}_\perp \psi$. Suppose that the changes in the shape of the auroral arc are realized through changes in the variables ($\omega$, $j_\parallel$) and that $E_\parallel$ and its electron acceleration are dropped. 

Ionospheric plasma motion including density waves is described by the two fluid equations. Considering the current dynamo layer (height of 100--150 km), we can assume that ions and electrons respectively yield the Pedersen drift $\mbox{\boldmath $v$}_{\rm i} = \mu_{\rm P} \mbox{\boldmath $E$}-D \mbox{\boldmath $\nabla $}_\perp \ln n_{\rm i}$ and the Hall drift $\mbox{\boldmath $v$}_{\rm e} = \mu_{\rm H} \mbox{\boldmath $E$} \times \mbox{\boldmath $B$}_0/B_0 - \mbox{\boldmath $j$}_\parallel/e n_{\rm e}$, with $\mu_{\rm P,H}$: mobilities and $D$: molecular diffusion coefficient. By integrating the continuity equations over the dynamo layer, the equations at $s=0$ become (see HW [2011] for details) 
\begin{eqnarray}
 \displaystyle && \partial_t n_{\rm e} + \mbox{\boldmath $v$}_\perp \cdot \mbox{\boldmath $\nabla$}_\perp n_{\rm e} = j_\parallel - R n_{\rm e} \\
 && \mbox{\boldmath $\nabla$}_\perp \cdot (n_{\rm e} \mu_{\rm P} \mbox{\boldmath $E$}) - \mbox{\boldmath $v$}_\perp \cdot \mbox{\boldmath $\nabla$}_\perp n_{\rm e} = D \nabla_\perp^2 n_{\rm e} - j_\parallel. 
\end{eqnarray}
Here, $R n_{\rm e}$ is a linearized recombination term, and the Hall mobility is normalized to be unity. We assume that $j_\parallel$ is carried by thermal electrons. Equation (4) includes the nonlinearity of the Pedersen and Hall current divergences. 

We used the 4th-order central difference method in space and 4th-order Runge-Kutta-Gill method in time to solve Eqs.\ (1)--(4). The number of grids were (256, 256, 128) for the $\nu$, $\varphi$, and $s$ directions, respectively. The time resolution was changed in accord with the Courant condition: $\max (\mbox{\boldmath $v$}_1 / \Delta x(s)) \Delta t < 0.25$. The numerical viscosity $\nu_{\rm v}$ and resistivity $\eta$ equaled $1\times 10^{-7} /B_0(s)$. Regarding the calculation domain $\mbox{\boldmath $x$}_\perp(s=0) \equiv [x,y]$, $x$ and $y$ pointed southward and eastward, respectively, in the southern hemisphere. We set a periodic boundary in the $\mbox{\boldmath $x$}_\perp$ direction, e.g., at $x$, $y=0$ and $l_\perp = 70$ km (thus $\Delta x \approx 0.27$ km) at the ionosphere $s=0$; this is valid since we take a local flux tube approximation ($L={\rm const}$). An asymmetric boundary for the magnetic field $\psi = 0$ (or $j_\parallel = 0$) was set at the magnetic equator $s=l$. At the ionospheric boundary of $\phi$, Eq.\ (4) was solved using the multigrid-BiCGStab method. 

For characteristic scales, the Alfv$\acute{\rm e}$n velocity and transit time are set to be $v_{\rm A}'\approx 1.5 \times 10^3$ km/s and $\tau_{\rm A}=l/v_{\rm A}' \approx 47$ s, while $l_\perp \approx 70$ km, the magnetic field $B_0 \approx 5.7 \times 10^{-5}$ T, and the electron density $n' \approx 3.8 \times 10^3$ cm$^{-3}$ are values at the ionosphere $s=0$. The drift velocity is $v_\perp' = v_{\rm A}' l_\perp/l \approx 1.5$ km/s. We set $v_{\rm A} = v_{\rm A}'$ along $s$ by using the dipole field $B_0(s)$ and a density profile $n_0(s)$. The ambient density at $s=0$ is set to be $n_0 = 10n'$; note that the above $v_{\rm A}'$ was determined using the F layer density ($\approx 7 \times 10^5$ cm$^{-3}$) and does not necessarily match this $n_0$. The other values are the same as in Hiraki [2013]: $\mu_{\rm P}/\mu_{\rm H} = 0.5$, $\Sigma_{\rm P}/\Sigma_{\rm A} = n_0\mu_{\rm P} = 5$, $D=4\times 10^5$ m$^2$/s, and $R = 2 \times 10^{-3}$ /s. 

We solved a linearized set of Eqs.\ (1)--(4) to determine the eigenfunctions ($\tilde{\phi}(s)$, $\tilde{\psi}(s)$, $\tilde{n}_{\rm e}(0)$) and frequency $\Omega$ of Alfven waves as functions of the perpendicular wavenumber $\mbox{\boldmath $k$}_\perp$ and the field-line harmonic number. For the above setting, we found that the fundamental mode with a frequency range of $\Omega \tau_{\rm A} \approx \frac{\pi}{2}$--$\pi$ has the maximum growth rate $\gamma \equiv {\rm Im} (\Omega) \tau_{\rm A}/\pi$. By fixing $k_x / 2 \pi = 1$ (hereafter, $k_x = 1$) that matches the arc form, the modes with $\gamma$ switch from $(k_x, k_y) = (1, 5)$ at $E_0 = 20$ mV/m to $(1, 2)$ at 80 mV/m as shown in Fig.\ 1. Here, the convection electric field $E_0$ is assumed to point poleward, so that the Pedersen current also points poleward ($x$), the Hall current eastward ($y$). We will study the $E_0$ dependence for the case of $(k_x, k_y) = (1,5)$ at Sec.\ 3. We will discuss the case of $(k_x, k_y) = (1,2)$ at Sec.\ 4 as well as (1,0), (2,0), and (3,0) modes. Also, diffusion and recombination in Eqs.\ (3) and (4) have an effective role in reducing the growth rate; see HW [2011] for details. 

\begin{figure}[b]
\includegraphics[width=1.0\columnwidth, bb=0 0 360 252, clip]{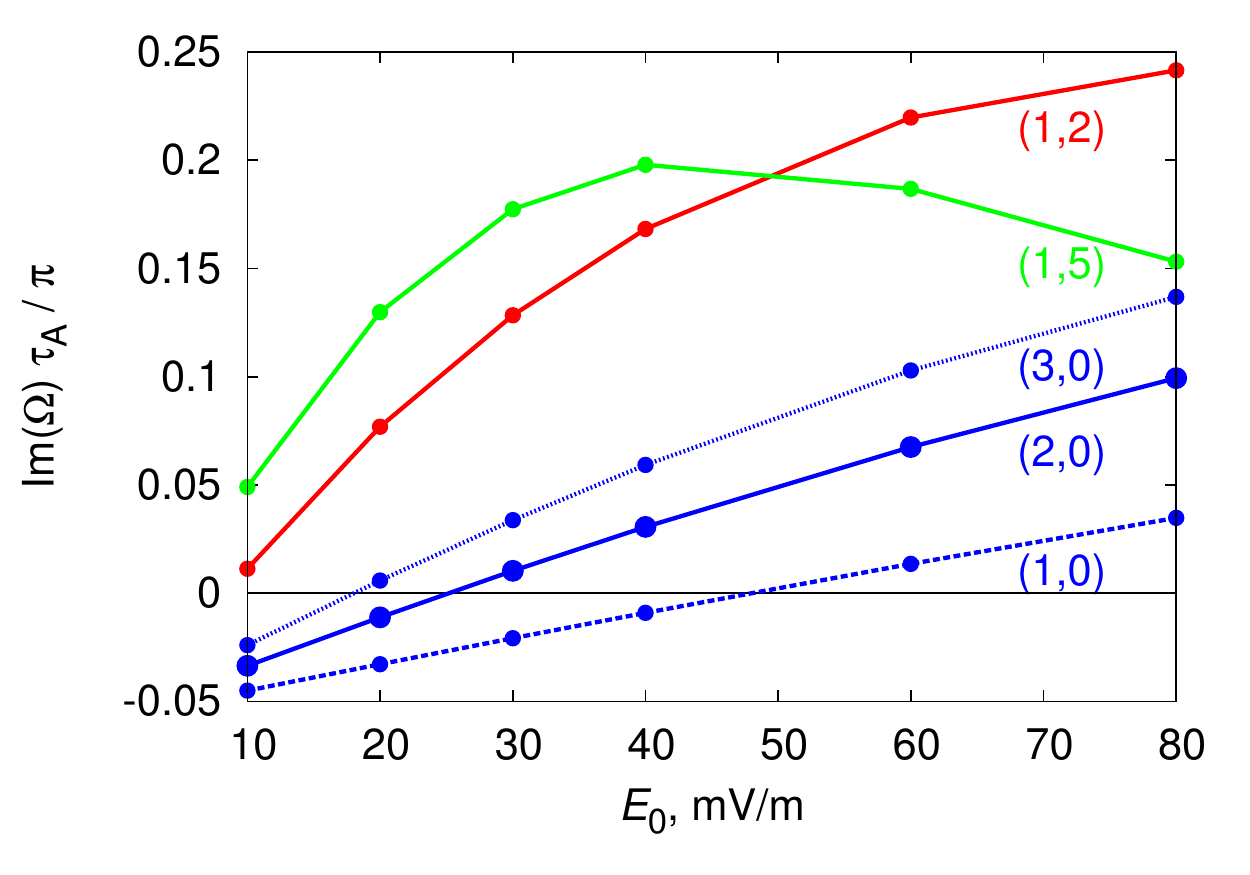}
\caption{Growth rates $\gamma\equiv {\rm Im}(\Omega) \tau_{\rm A} / \pi$ of several modes $(k_x, k_y) = (1,2)$, (1,5), (1,0), (2,0), and (3,0) as functions of the convection electric field $E_0$; $\gamma = 0.1$ corresponds to the time scale of $\tau\approx 2.5$ min.}
\end{figure}

We performed a 3D simulation to ascertain the growth of feedback eigenmodes ($\tilde{\phi}$, $\tilde{\psi}$, $\tilde{n}_{\rm e}$), from an initially east-west aligned auroral arc, in the poleward convection field $E_0$. Here, the perturbed potentials are partitioned into the arc component and the feedback eigenmode with $\mbox{\boldmath $k$}_\perp$ shown above, as $(\phi, \psi, n_{\rm e}) = (\phi_{\rm a}, \psi_{\rm a}, n_{\rm ea}) + (\tilde{\phi}, \tilde{\psi}, \tilde{n}_{\rm e})$. The arc potential is yielded as the fundamental wave form of $\phi_{\rm a}(s) \propto \frac{1}{B_0(s)} \sin(\frac{\pi}{2l} s)$ while $\psi_{\rm a}(s) = n_{\rm ea} = 0$ for simplicity. The essence of our results were unchanged for the choice of $\psi_{\rm a}$ and $n_{\rm ea}$ since these quickly adjust to $\phi_{\rm a}$. The perpendicular function of $\phi_{\rm a}$ is assumed to be Gaussian-like with $l_{\rm a}\approx10$ km and $E_{\rm a}\approx 20$ mV/m at $s=0$ (see Fig.\ 2); note that negative $\omega$ is quickly produced. The electric field points equatorward at the poleward edge and is reversed at the equatorward side. It is accompanied with a counter-clockwise flow shear across the arc, though it is too weak to trigger some instability. The feedback eigenmode has an amplitude of $|\tilde{\phi}| = |\tilde{\psi}| = 10^{-4}|\phi_{\rm a}|$ at $t=0$. Application of the above setting to the observed arc deformation, ignoring field-aligned electron acceleration and related source process, is discussed at Sec.\ 4.

\section{Results}
Figure 2 shows the temporal variation in vorticity $\omega$ ($t/\tau_{\rm A} = 0.1$, 4, and 7) at the ionosphere $s=0$, in the case of the feedback mode $(k_x, k_y) = (1,5)$ and the poleward field $E_0=60$ mV/m. Note that since the results are shown in the frame of a convection drift $v_0$, structures mainly move westward in the rest frame. As density waves related to the Pedersen current $j_{\rm P} \parallel x$ propagate poleward, a new arc is produced (by splitting) through a current divergence between the wave and the initial arc. Once these arcs become dark, a vortex street forms in the poleward brightening arc at $t \approx 3$ min (panel (b)). Another vortex street forms in the equatorward arc during the next $\approx 2$ min; it coalesces with the poleward one into larger vortices (30--40 km). The amplitude of the flow is $\max |v_1| \approx 0.32$ km/s in panel (b) and grows to $\approx 0.75$ km/s. In this case, the convection flow is $v_0 \approx 1.1$ km/s. It is also clear that in panel (b) the flow is counterclockwise ($\parallel j_{\rm H} \sim -v_{0y}$) at the poleward edge of the vortices and at the front of the $j_{\rm P}$-density waves. On the other hand, in panel (c), a clockwise flow is added by negative $\omega$. It is clear that the vortex street can be produced from an arc under a high $E_0$. 

\begin{figure}
\includegraphics[scale=0.75, bb=120 80 400 720, clip]{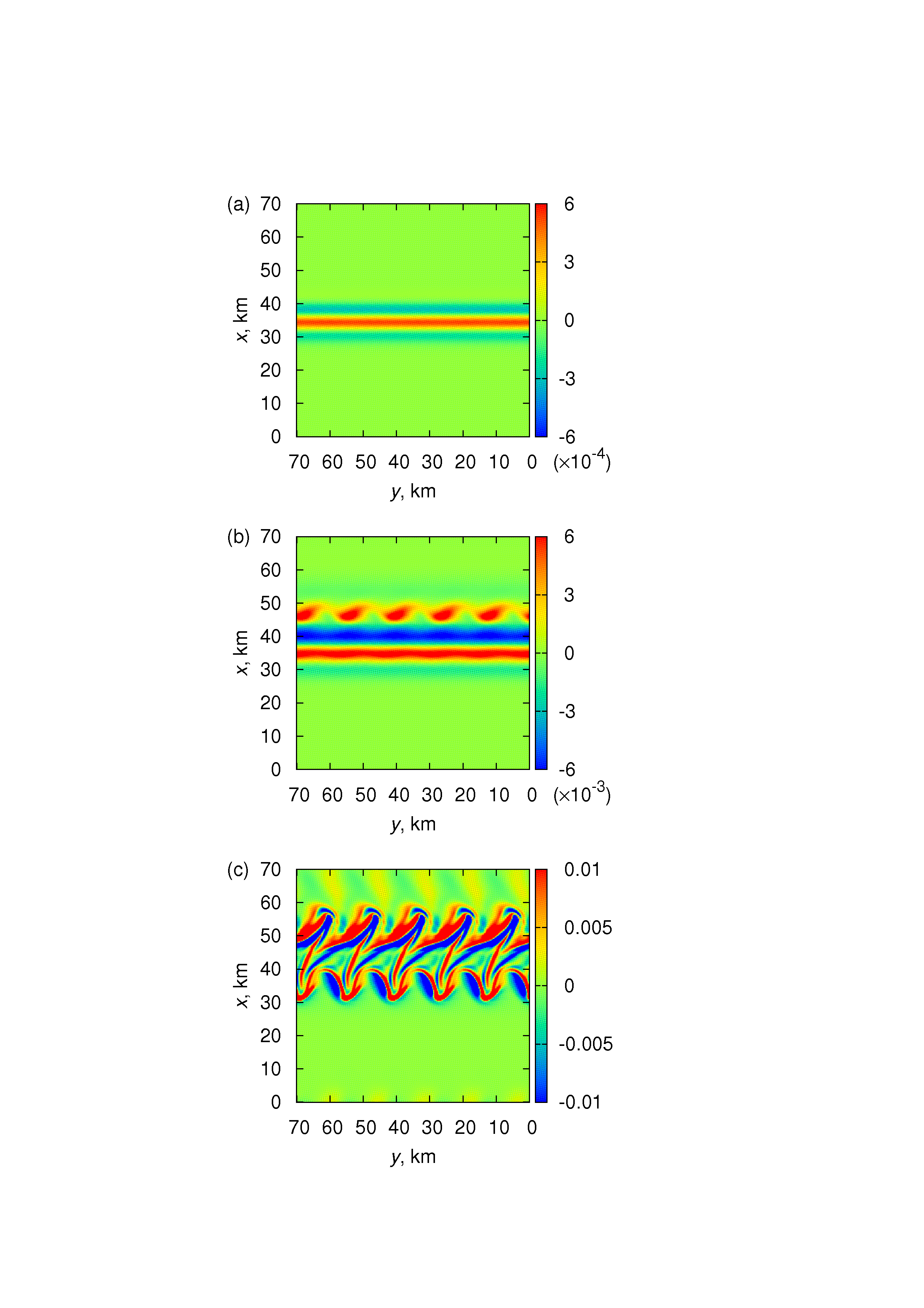}
\caption{Vorticity $\omega$ at the ionosphere $s=0$ at the time of (a) $t/\tau_{\rm A}=0.1$, (b) 4, and (c) 7, in the case of a convection electric field $E_0 = 60$ mV/m. Note that the $y$-axis is reversed ($x \times y \parallel b_0$) because we are considering the southern hemisphere.}
\end{figure}

Figure 3 shows the temporal variation in $j_\parallel$ (a--c) and $n_{\rm e}$ (d--f) at $s=0$ accompanied by $\omega$ in Fig.\ 2. The amplitude of the current is $\max (j_\parallel > 0) \approx 3.1$ $\mu$A/m$^2$ in panel (b) and grows to $\approx 30$ $\mu$A/m$^2$ in panel (c) [c.f., Haerendel, 2010]. We can easily find that $j_\parallel$ is almost out of phase with $\omega$ in the linear stage of feedback instability. Downward currents $j_\parallel <0$ are produced just on the vortex street at $t/\tau_{\rm A}=4$, connecting to the equatorward Pedersen current, and upward currents are induced at the equatorward arc. The vortex street with $j_\parallel<0$ (electron loss) moves poleward due to the background Pedersen drift, and the electron density decreases just at its back ($x\approx 42$ km). On the other hand, $n_{\rm e}$ increases at the equatorward side of $j_\parallel>0$ (electron supply). Structures of $j_\parallel$ are complicated at the nonlinear stage of $t/\tau_{\rm A}=7$, but intense upward $j_\parallel$ (up to 30 $\mu$A/m$^2$) appear on both sides of the poleward edge of vortices; e.g., at ($x$, $y$) = (55 km, 10 km). The electron density is depleted up to $-30 \%$ due to the net $j_\parallel <0$ at the poleward side, while a tongue-like structure of density enhancement forms at the equatorward side. 

\begin{figure*}
\includegraphics[scale=0.75, bb=100 80 700 720, clip]{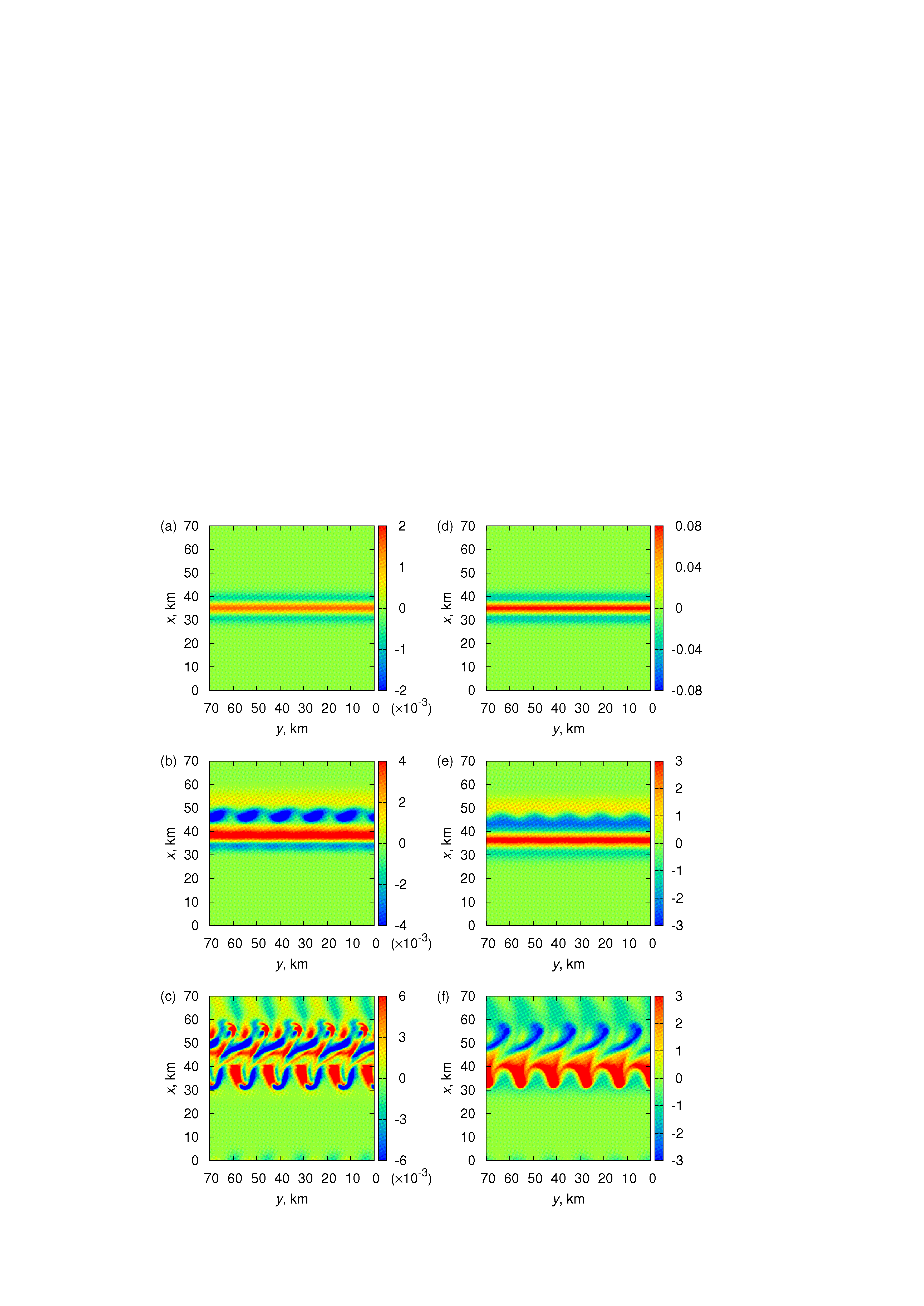}
\caption{Same plots as Fig.\ 2 but for (a)--(c) field-aligned current $j_\parallel$ and (d)--(f) electron density $n_{\rm e}$ at the ionosphere $s=0$. The unit of $j_\parallel$ is normalized by 650 $\mu$A/m$^2$, i.e., $10^{-3}$ equals 0.65 $\mu$A/m$^2$. See text for $n_{\rm e}$.}
\end{figure*}

Figure 4 shows the field line distribution of the average vorticity $\langle \omega \rangle(s)$ during certain periods along with its cross section at $s = l$; $\max |v_1| \approx 31$ km/s at $t/\tau_{\rm A} = 7$. The Alfv$\acute{\rm e}$n wave propagates to the ionosphere at $t/\tau_{\rm A} = 1$ but goes away at $t/\tau_{\rm A} = 2.5$, which means the maximum $\langle \omega \rangle$ at $s=l$. Some waves still remain at $s=0$. Waves come back again to $s=0$ by $t/\tau_{\rm A} = 4$. We suppose that the apparent wave propagation time becomes longer ($> 1$) because the initial function is deformed, which means generation of a new wave on the way. The amplitude at $s=4$--9 R$_{\rm E}$ decreases while the vortex street forms during this period. Waves return to the magnetosphere during $t/\tau_{\rm A} = 4$--5, and the amplitudes in the region of $s>3$ R$_{\rm E}$ increase. Although partial reflections continuously occur, the waves (or $\langle \omega \rangle_{\max}$) are on the $s=0$ side at $t/\tau_{\rm A} = 6$ and on the $s=l$ side at $t/\tau_{\rm A} = 7$. The nonlinear coupling to the fast density waves (large $E_0$) causes a rapid growth of Alfv$\acute{\rm e}$n waves through slippage and partial reflection, resulting in a pileup of vortices. It is also clear that the vortex pattern changes between $s=0$ and $l$; an equi-potential mapping cannot be inferred at this scale of aurora. We further see that radially inward weak flows develop at $x \approx 1700$ km and $\approx 3000$ km through the effect of arc splitting.

\begin{figure}
\includegraphics[scale=0.7, bb=90 80 650 650, clip]{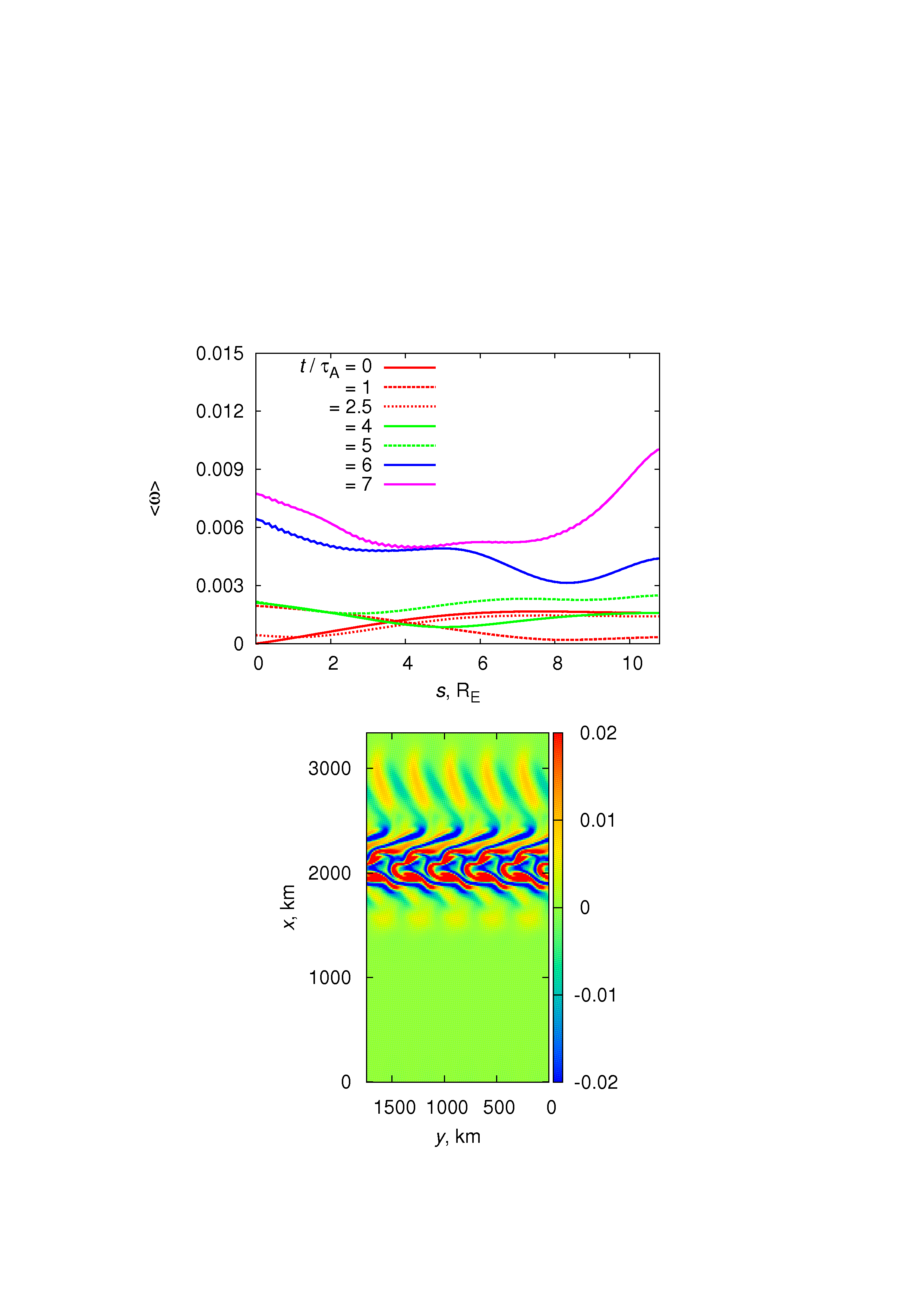}
\caption{Upper: Root-mean-square vorticity $\langle \omega \rangle$ in $\mbox{\boldmath $x$}_\perp(s)$ as a function of $s$ and $t$ in the case of Fig.\ 2. Lower: Vorticity $\omega$ at the magnetic equator $s=l$ at $t/\tau_{\rm A}=7$.}
\end{figure}

Figure 5 shows the average electron density $\langle n_{\rm e} \rangle(t)$ for $E_0 = 20$--80 mV/m. When the vortex street forms (see Fig.\ 2), it increases by up to 10--15 $\%$ of $n_0 = 10 n'$. In the case of $E_0=20$ mV/m, arc splitting occurs just before every minimum of $\langle n_{\rm e} \rangle$, and new arcs repeatedly vanish. Although shears of these arcs prevent the growth of the mode $(k_x, k_y) = (1,5)$, eventually a north-south elongated structure (not vortex street) grows at $t/\tau_{\rm A} \approx 27$. The knowledge we get here is the change in the growth pattern between $E_0 = 20$--40 mV/m indicates the existence of a critical field $E_{\rm cr}$ for deformation of the arc into the vortex street. Further sensitivity studies (not shown) confirmed that the results related to $E_{\rm cr}$ are independent of the initial conditions of arc itself, i.e., width $l_{\rm a}$ of 10--20 km, field $E_{\rm a}$ of 20--80 mV/m, and polarity ${\rm sgn}(\phi_{\rm a})$. 

\begin{figure}[b]
\includegraphics[width=1.0\columnwidth, bb=0 0 360 252, clip]{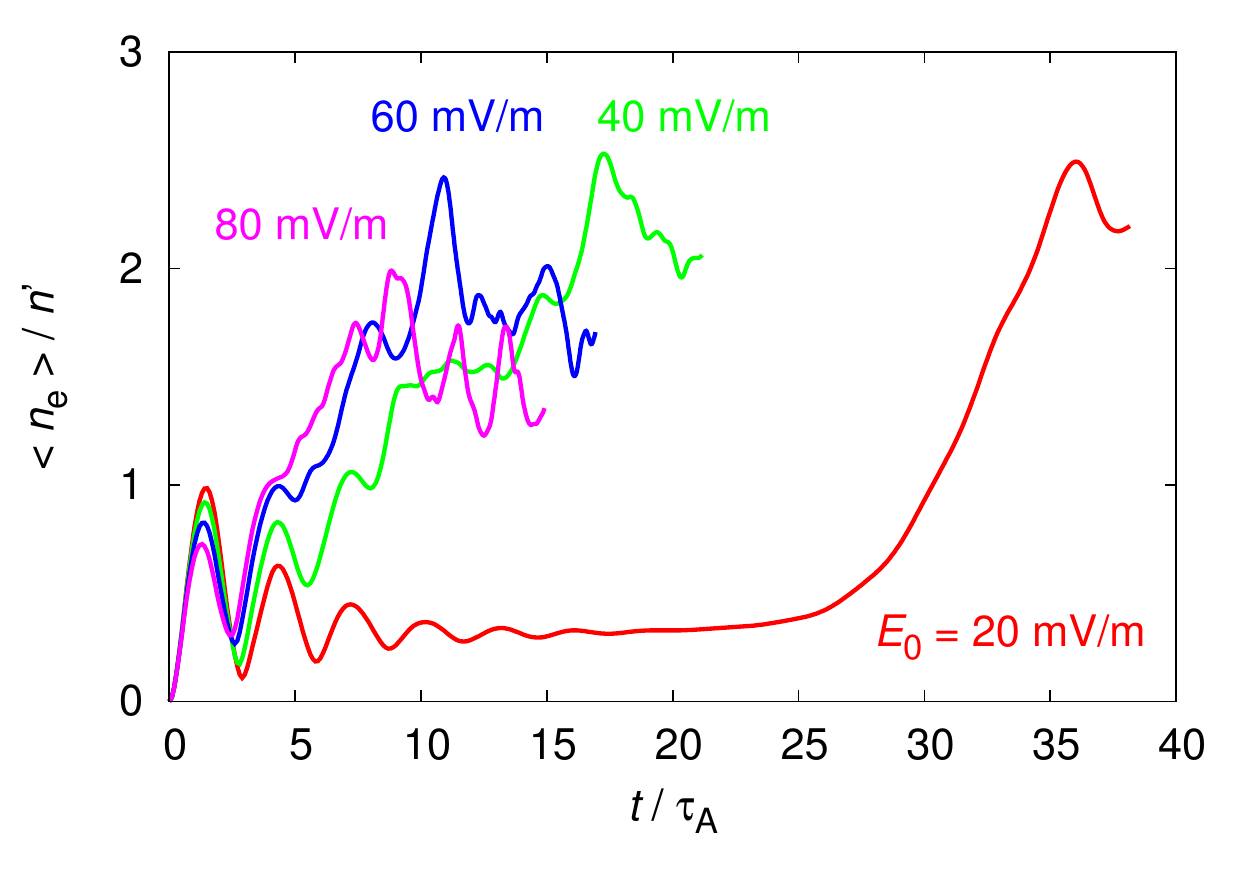}
\caption{Time variation of root-mean-square electron density $\langle n_{\rm e} \rangle$ at the ionosphere for $E_0=20$, 40, 60, and 80 mV/m.}
\end{figure}

\section{Discussion}
Let us first mention an overarching problem with regard to the application of our modeled system to the real world where auroral mechanisms, i.e., field-aligned electron acceleration and the induced ionospheric density change, are at work from the first brightening of the onset arc. We suppose that these are generally important in the arc system but do not directly contribute to the arc deformation itself. The arc deformation could start from destabilization of the Alfv$\acute{\rm e}$n waves, and field-aligned currents are amplified to be 4--20 $\mu$A/m$^2$ in their nonlinear stage as shown in Fig.\ 3. The high current density can yield a strong field-aligned voltage if some resistivity as electron inertia and the realistic deep cavities are included in the system. As our first step by a minimal model setup without these features, we addressed the pure nonlinear coupling between an arbitrarily placed arc structure and Alfv$\acute{\rm e}$n eigenmodes. At the next step, we will investigate the generation of field-aligned electric fields, the related electron acceleration and ionospheric source processes in our nonlinear MI feedback system. 

We would further provide two notes before analyzing our calculation results to compare with observations. The present study ignoring field-aligned potentials regards a pattern of vorticity $\omega$ as that of auroral arc. This treatment is supported by imager and radar simultaneous observations [Hosokawa et al., 2013], where the onset arc accompanies a counter-clockwise flow shear. It implies that the auroral arc strongly couples to the flow field (or vorticity) at the ionosphere $s=0$. Since the assumption was made that field-aligned currents are carried only by thermal electrons at $s=0$, regions of the upward current $j_\parallel >0$ at $s=0$ does not necessarily represent the auroral luminosity. The associated electrons only flow up magnetic fluxes but cannot form the auroral luminosity. The next study should address on the relationship between $\omega$, $j_\parallel$ at $s=0$, those at the region of high $E_\parallel$, and the related electron acceleration. 

The second note is on our initial setup of the auroral arc. Although the system in the case of Fig.\ 2 is highly unstable with the imposed arc, the system under a weak convection electric field (e.g., $E_0 = 20$ mV/m) is demonstrated to be stable as in Fig.\ 5 and discussions below. Destabilization of the arc along with feedback eigenmodes is not self-evident, and thus we consider our setup be natural. The scenario is that the arc appears long time (30--60 min) before substorm onset under a weak (westward) $E_0$, and the source is at the magnetospheric end. After the poleward component of $E_0$ is enhanced up to a critical level (see the next paragraph), the arc rapidly grows as shown in Figs.\ 2 and 3. 

Let us describe the changes in the wave growth patterns in the convection electric field of $E_0 = 20$--40 mV/m (Fig.\ 5) in the context of arc splitting. The behavior can be understood by the linear growth rate of feedback eigenmodes in Fig.\ 1 because they certainly grow in spite of oscillatory motion of $\langle n_{\rm e} \rangle$ through Alfv$\acute{\rm e}$n wave propagation (see also Fig.\ 4). Figure 1 shows that $\gamma$ of the (2,0) mode switches from negative to positive at 20--40 mV/m. On the other hand, $\gamma$ of the (1,5) mode increases by $\approx 1.5$ times in this range, but decreases in the higher regime; as these tendencies are inconsistent with the above growth pattern, this mode is not considered to be the main trigger. $\gamma$ of the (1,2) mode monotonously increases, which affects its saturation level as shown below, but it is not directly related to the change. Consequently, we can infer that the vortex street grows just after the arc splits to generate the (2,0) mode. Alfv$\acute{\rm e}$n wave amplification follows the growth of the mode and leads to a cumulative growth of the (2,0) and (1,5) modes, forming into the vortex street. The critical field $E_{\rm cr}$ for the southward modes ($k_y=0$, or the (2,0) mode in this case) is $\approx25$ mV/m.

An increase in $\langle n_{\rm e} \rangle$ in the case of $E_0 = 20$ mV/m (Fig.\ 5) means that eastward modes ($k_y \ne 0$) can grow, though it takes long time ($\approx 25$ min), after a couple of arc splittings; remember the mode is defined in the westward moving frame of $v_0$. However, these modes are decoupled from the southward modes during this growth, which is not consistent with observation of the arc changing into the vortex street [e.g., Sakaguchi et al., 2009]. We conclude that this case is a product of certain ideal conditions, and the critical field for vortex formation is the value given in the previous paragraph. 

Let us see to what extent the vortex street can brighten. Figure 6 shows the results for the feedback mode $(k_x, k_y) = (1,2)$ and $E_0 = 60$ mV/m. As in the case of $(k_x, k_y) = (1,5)$ (Fig.\ 2), the initial arc repeatedly splits until $t / \tau_{\rm A} \approx 5$, brightens, and deforms into a twin vortex after that. Note that in every case of (1,2) with $E_0 \ge 40$ mV/m, the saturation level of $\langle n_{\rm e} \rangle$ shifts to a higher value, up to $\approx 38 \%$ from $n_0$, than the level of the (1,5) cases. This behavior is not expected from the linear growth rates, i.e., $\gamma(1,2) < \gamma(1,5)$ at $E_0 > 40$ mV/m, depicted in Fig.\ 1. We could claim that the (1,2) mode is well-suited for nonlinear coupling to the (2,0) mode. What can be inferred from this result is that the extent of arc intensification related to vortices is controlled by the growing eastward feedback eigenmodes. The modes were mainly determined by conductances $\mu_{\rm P}/\mu_{\rm H}$ and $\Sigma_{\rm P}/\Sigma_{\rm A}$ along with $E_0$. 

\begin{figure}
\includegraphics[scale=0.7, bb=90 80 650 530, clip]{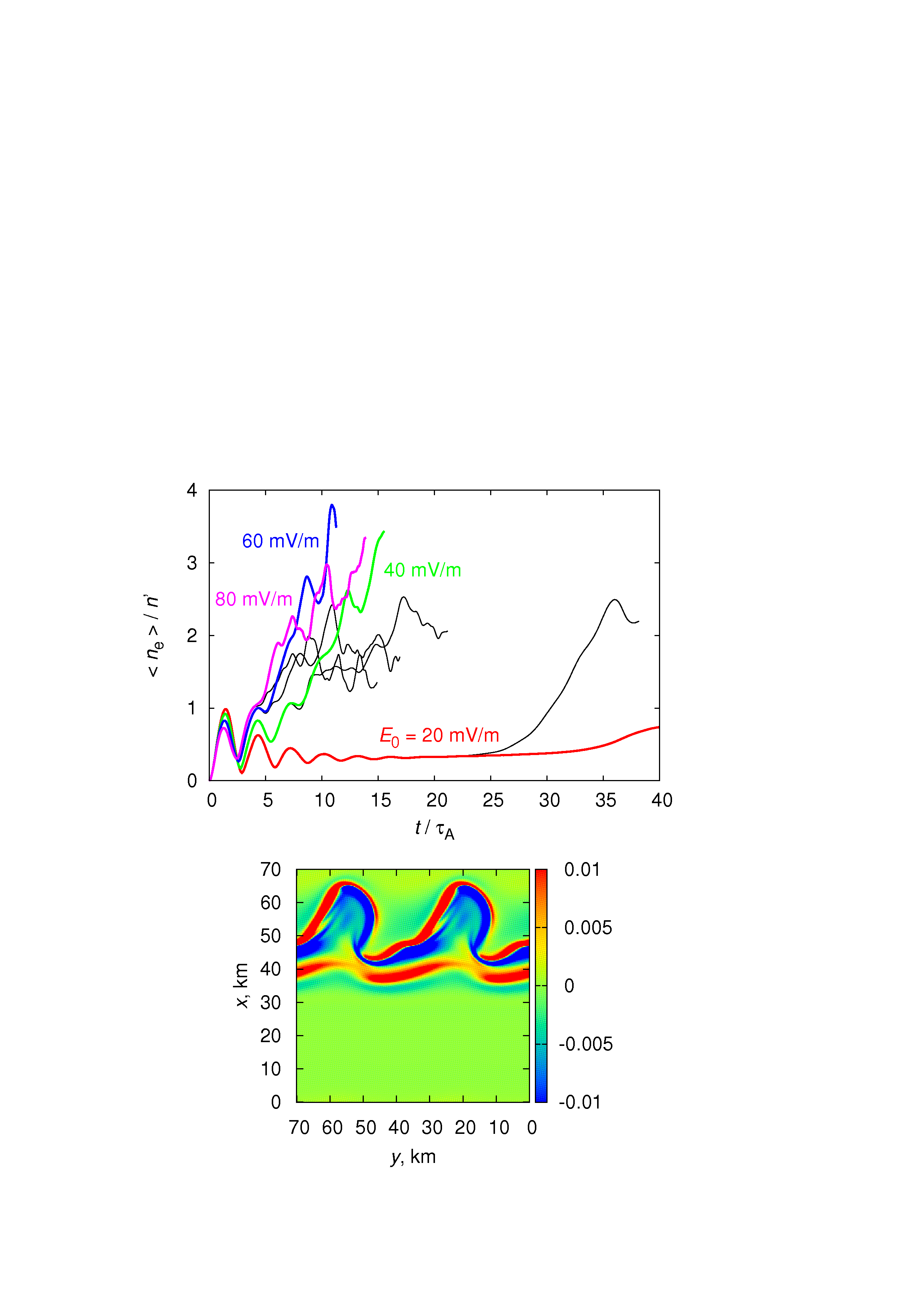}
\caption{Upper: Time variation of $\langle n_{\rm e} \rangle$ at the ionosphere $s=0$ for $E_0=20$, 40, 60, and 80 mV/m in the case of $(k_x, k_y) = (1,2)$; black lines are those shown in Fig.\ 5. Lower: Vorticity $\omega(s=0)$ at $t/\tau_{\rm A}=7.6$ for $E_0 = 60$ mV/m.}
\end{figure}

Let us compare our results with the observed features of arcs during substorm onset. Ground-based optical observations have indicated that the arcs often flap and split just before vortex street formation [e.g., Motoba et al., 2012; Private comm.\ with K.\ Hosokawa and K.\ Sakaguchi]. These behaviors imply that ionospheric currents and density waves are increased by their feedback coupling to Alfv$\acute{\rm e}$nic $j_\parallel$. The vortex sizes produced by a feedback growth of the (1,5) and (1,2) modes are $\approx14$ km and 35 km, respectively. We should notice that our simulation domain is a dipole field line ($L \approx 8.5$), when the vortex sizes are compared to those seen in the onset arc as 40--80 km [Sakaguchi et al., 2009; and references therein]. The field line length $l$ can be extended if an actual magnetotail field line is considered. Then, $l_\perp \propto l$ is also extended, and the above difference (factors of 2--3) between the vortex size in our model results and that in observations is not up to date the matter for explaining the behavior of the onset arc. Another point is on the dynamic range of optical (all-sky camera) observations. In case of the above paper, power spectra in scales of $< 40$ km are noisy, and further the scale estimation strongly depends on the zenith angle of arcs. We urge high resolution measurements of the onset arc to estimate its real wavelength; multi-point stereo observations are anticipated. 

The next point is on the magnitude of $E_0$. The critical field $E_{\rm cr} \approx 25$ mV/m (i.e., a flow of 0.43 km/s) we present here fairly agrees with the enhanced convection flows ($\ge0.5$ km/s) observed before onset [Bristow and Jensen, 2007]. Further comparison provides a useful restriction for interpreting statistical data, as well as the direction of $E_0$ assumed to be poleward in this paper. We urge that detailed measurements of pre-conditioning of onset arcs should be undertaken in the future. Moreover, there is another problem related to the growth time scale. The scale of the observations, $\tau = 1$--2 min [e.g., Mende et al., 2009] is shorter than that of our calculation, $\tau = 3$--5 min. The Alfv$\acute{\rm e}$n transit time we use is as large as $\tau_{\rm A}\approx 47$ s, but the growth time was shown to be shorter in more realistic cases ($\tau_{\rm A}\approx 25$ s) with ionospheric $v_{\rm A}$ cavities [HW, 2012]. Therefore, simulations in the cases can explain the small $\tau$ in the observations. 

The results in this paper indicate that feedback modes are prevented from growing by the initial arc when $E_0 < E_{\rm cr}$, but they grow together into a vortex street when $E_{\rm cr} < E_0$. The vortex street is a manifestation of the nonlinear evolution of Alfv$\acute{\rm e}$n waves in the MI coupling system. The problems remaining to be solved are i) how are waves trapped in the ionospheric cavity region to form a field-aligned electric field when the field-aligned current is amplified? And ii) why does the active arc expand poleward in the next step? The second question is closely related to coupling with magnetotail high-$\beta$ plasma dynamics [e.g., Henderson, 2009].

\section{Conclusion}
By performing three-dimensional nonlinear MHD simulations including Alfv$\acute{\rm e}$n eigenmode perturbations most unstable to the ionospheric feedback effects, we reproduced the auroral vortex street that often appears just before substorm onset. We found that i) the initial arc splits, intensifies, and deforms into a vortex street within $\approx5$ min, ii) growth pattern changes at a convection electric field of $E_{\rm cr}=25$ mV/m, and iii) the extent of arc intensification is controlled by the nonlinear behavior of feedback eigenmodes with $k_y\ne0$. The results of our simulation indicate that the vortex street is a consequence of coupling between the shear Alfv$\acute{\rm e}$n waves carrying field-aligned currents and the ionospheric density waves driven by the Pedersen/Hall currents.


\newpage 

\end{document}